УДК 371

**А.М. СТРЮК,**
*кандидат педагогічних наук,*
*доцент кафедри моделювання та програмного забезпечення*
*ДВНЗ «Криворізький національний університет»*

**С.О. СЕМЕРІКОВ,**
*доктор педагогічних наук, професор,*
*в. о. завідувача кафедри фундаментальних дисциплін*
*ДВНЗ «Криворізький національний університет»*

# МОДЕЛІ КОМБІНОВАНОГО НАВЧАННЯ

У статті на основі існуючих моделей навчання презентовано авторську організаційну модель комбінованого навчання у ВНЗ, що передбачає використання системи управління навчанням та відображає поточний стан розвитку теорії та методики використання ІКТ в освіті.

*Ключові слова: модель, комбіноване навчання, організаційна модель комбінованого навчання.*

**Актуальність проблеми.** Сучасна система освіти все більше набуває якостей мобільності та відкритості: розвиток комунікацій розмиває кордони між державами та глобалізує ринок праці за рахунок підвищення соціальної мобільності; уніфікація систем освіти різних країн, зумовлена зростаючою потребою у підготовці фахівців для глобалізованого світу, потребує підвищення навчальної мобільності; зростання соціальних стандартів вимагає широкої інклюзії осіб з особливими потребами у навчальний процес та виробничу діяльність; швидкість зміни змістового наповнення навчальних дисциплін пов'язана з високим темпом модернізації виробничих технологій і потребує переходу від старої парадигми «навчання на все життя» до нової – «навчання протягом усього життя» та забезпечення професійної мобільності; поширення концепції Open Source з програмного забезпечення на навчальні матеріали приводить до виникнення відкритих, вільно поширюваних навчальних курсів.

Кейптаунська декларація відкритої освіти «Відкриваючи майбутнє відкритим освітнім ресурсам» [15], прийнята у 2007 р., визначає, що «рух відкритої освіти ... ґрунтується на засадах того, що кожен без застережень повинен мати свободу використовувати, адаптувати, поліпшувати та поширювати ... навчальні матеріали, ліцензовані відкритими ліцензіями, ... підручники, ... програмне забезпечення та інші матеріали, які допомагають вчити та навчатися... [і] розвивають ... культуру навчання, творення, обміну і співпраці у швидкозмінному суспільстві знань». При цьому підкреслюється, що «відкрита освіта не вичерпується лише відкритими освітніми ресурсами. Вона виростає з відкритих технологій, що сприяють співпраці та гнучким підходам до навчання». У рішеннях Всесвітньої конференції ЮНЕСКО з вищої освіти 2009 р. вказується, що формування компетентностей XXI ст. можливе при комплексному застосуванні відкритої та дистанційної освіти і засобів ІКТ, що створюють умови широкого доступу до якісної освіти, зокрема – на основі відкритих освітніх ресурсів [1; 3].

Технологічною основою вказаних тенденцій є сучасні ІКТ навчання, серед яких провідне місце посідають технології електронного, дистанційного та мобільного навчання. **Відкритість освіти** насамперед передбачає використання всіма суб'єктами навчання таких засобів ІКТ, що надають можливість вільного доступу до навчальних матеріалів і освіти в цілому. У дослідженнях В.Ю. Бикова показано, що застосування ІКТ для реалізації відкритої освіти сприяє навчальній та професійній мобільності, індивідуалізації освітніх траєкторій, реалізації інклюзивної та андрагогічної освіти. ІКТ мережевого навчання мають забезпечувати відкритий доступ не лише до традиційних навчальних матеріалів у вигляді навчальних посібників, підручників тощо, а й до навчального лабораторного обладнання як безпосередньо через віддалене управління, так і опосередковано через застосування віртуальних лабораторій. Аналіз сучасних засобів ІКТ відкритої освіти показав, що найуніверсальнішими серед них є відкриті системи управління навчанням, спільними властивостями яких є: відкритість програмного коду та процесу розробки; апаратна та програмна мобільність; підтримка педагогічних технологій електронного, дистанційного та мобільного навчання.

**Аналіз досліджень і публікацій.** Застосування відкритих систем управління навчанням створює умови для надання навчальному процесу якості неперервності шляхом технологічної інтеграції аудиторної та позааудиторної роботи в систему комбінованого навчання, різні аспекти якого розглядали Дж. Берсін, В.Ю. Гнезділов, П. Джонс, Р. Кертіс, Т.І. Коваль, А. Рейд-Янг, Є.М. Смирнова-Трибульська, Б.І. Шуневич та ін. У дослідженнях Н.В. Рашевської, А. Хейнце, С.В. Шокалюк підкреслюється, що найвища ефективність комбінованого навчання досягається тоді, коли засоби ІКТ комбінованого навчання виступають також як об'єкт вивчення: у середній школі – при навчанні інформатики, у вищій – при підготовці фахівців з інформаційних технологій, попит на яких не лише постійно зростає, а і є суспільно зумовленим.

**Мета статті**: розкрити сутність моделі комбінованого навчання у ВНЗ.

**Виклад основного матеріалу.** Досвід застосування синхронних (спільна спеціально організована навчальна діяльність у визначений час у визначеному місці) та асинхронних (індивідуальна навчальна діяльність, що має бути виконана за певний час) форм організації навчання у ВНЗ показує, що у навчальному процесі вони комбінуються: так, провідна синхронна форма – лекція – супроводжується синхронною (фронтальною), синхронно-асинхронною (груповою) та асинхронною (груповою) лабораторною роботою. При цьому найвищий ступінь асинхронності, а також найбільшу частину самостійної роботи традиційно мають дистанційне та мобільне навчання. Доцільність комбінування різних форм організації навчання зумовлена тим, що, з одного боку, впровадження технологій електронного, дистанційного та мобільного навчання в аудиторне навчання надає можливість комп'ютеризувати самостійну роботу, а з іншого – частка самостійної роботи у навчальному плані визначає вибір форми навчання з відповідною ІКТ-підтримкою.

Аналіз різних трактувань *комбінованого навчання* (Т.І. Коваль, Н.В. Рашевської, Ю.В. Триуса та інших дослідників) надав можливість визначити його як *цілеспрямований процес здобування знань, умінь та навичок в умовах інтеграції аудиторної та позааудиторної навчальної діяльності суб'єктів освітнього процесу на основі використання і взаємного доповнення технологій традиційного, електронного, дистанційного та мобільного навчання* [21]. У цьому визначенні підкреслюються проміжна роль комбінованого навчання між традиційним (переважно аудиторним) і дистанційним (переважно позааудиторним) навчаннями, провідна роль ІКТ в організації навчальної діяльності, що дає змогу розглядати ІКТ комбінованого навчання та його відповідність системним принципам відкритої освіти (мобільності учасників навчального процесу, рівного доступу до освітніх систем, надання якісної освіти, формування структури та реалізації освітніх послуг [14, 55–56]).

Під технологією навчання М.І. Жалдак та Ю.В. Триус розуміють сукупність тих компонентів методичної системи навчання, що відповідають на питання «як навчати?»: методи, засоби та форми організації навчання [23, 248]. За такого визначення застосування комбінованого навчання також є інноваційною технологією [20, 39]. Ю.В. Триус визначає інноваційну педагогічну технологію як систему оригінальних, новаторських способів, прийомів педагогічних дій і засобів, що охоплюють цілісний навчально-виховний процес від визначення його мети до очікуваних результатів, і які цілеспрямовано, систематично й послідовно впроваджуються в педагогічну практику з метою підвищення якості освіти [22]. Проек-тування нової освітньої технології, згідно з дослідженнями Н. В. Бордовської [20], передбачає такі дії: аналіз потреб освітньої практики в новій технології; розробка моделей такої технології; детальний опис її специфіки, ідентифікації з існуючими технологіями; визначення основних засобів та умов, що необхідні для реалізації розробленої моделі; оцінка ефективності нової технології порівняно з іншими технологіями.

Таким чином, використання комбінованого навчання в освітньому процесі потребує побудови моделі, адаптованої до умов його застосування.

Комбіноване навчання інтегрує синхронні та асинхронні комунікаційні технології, формальне та неформальне навчання, друковані та електронні навчальні матеріали, онлайнову та офлайнову фасилітацію, забезпечуючи умови для створення якісних інтерактивних навчальних матеріалів для самонавчання та неперервної підтримки процесу навчання [2]. Розробка моделі комбінованого навчання, на думку Б. Хана, вимагає урахування вимог: педагогічних, технологічних, інституційних, етичних, управлінських, ресурсних, інтерфейсних та оцінювальних [9, 4].

У праці Є.М. Смирнової-Трибульскої [19] проаналізовано стан дослідженості і розробленості концепції дистанційного навчання в теорії і практиці безперервної освіти, зокрема вищої і післядипломної, встановлена сукупність особливостей дистанційного навчання як одного з перспективних підходів до організації педагогічного процесу у вищих педагогічних навчальних закладах і в післядипломній освіті. Спроектована в дослідженні модель інтеграції очної та дистанційної форм для ВНЗ, наведена на рис. 1, поділяє форми організації навчання, відносячи, зокрема, лабораторні роботи лише до дистанційного навчання, а семінарські та практичні заняття – лише до традиційного [18, 360].

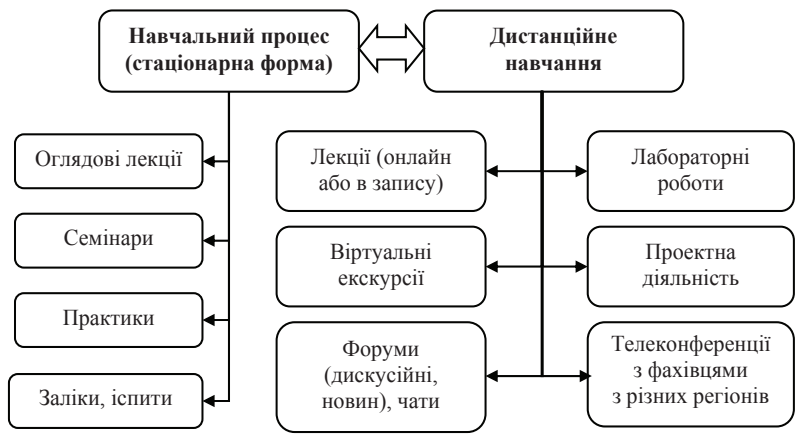

**Рис. 1. Модель інтеграції очної та дистанційної форм для ВНЗ (за Є.М. Смирновою-Трибульскою)**

Автори системи підтримки комбінованого навчання Networked Learning Ecology – North America (NLENA) [4] пропонують поєднувати форми організації аудиторного навчання, он-лайн-навчання та практичної підготовки, специфічні для мобільного навчання, що особливо актуально для технічних ВНЗ [17] (рис. 2). Трикомпонентну структуру має і модель комбінованого навчання корпорації Sealund (рис. 3), проте, на відміну від попередньої, вона включає в себе не лише форми, а й діяльність: технологію електронного навчання у поєднанні з моделюванням та ігровим підходом. Усі види діяльності консультативно підтримуються.

Дослідники німецької компанії Allconsulting GmbH [5] пропонують інше поєднання форм та методів комбінованого навчання у трикомпонентній моделі, що за структурою відповідає запропонованому нами тлумаченню комбінованого навчання (рис. 4). Однією з характеристик мобільного навчання автори вважають «швидке навчання» (rapid learning), яке, за Ч.М. Вебером, є основою для перепідготовки фахівців електронної промисловості на виробництві [13]. Так само, як і у попередній, у моделі Allconsulting GmbH провідними формами організації навчання є форми практичної підготовки.

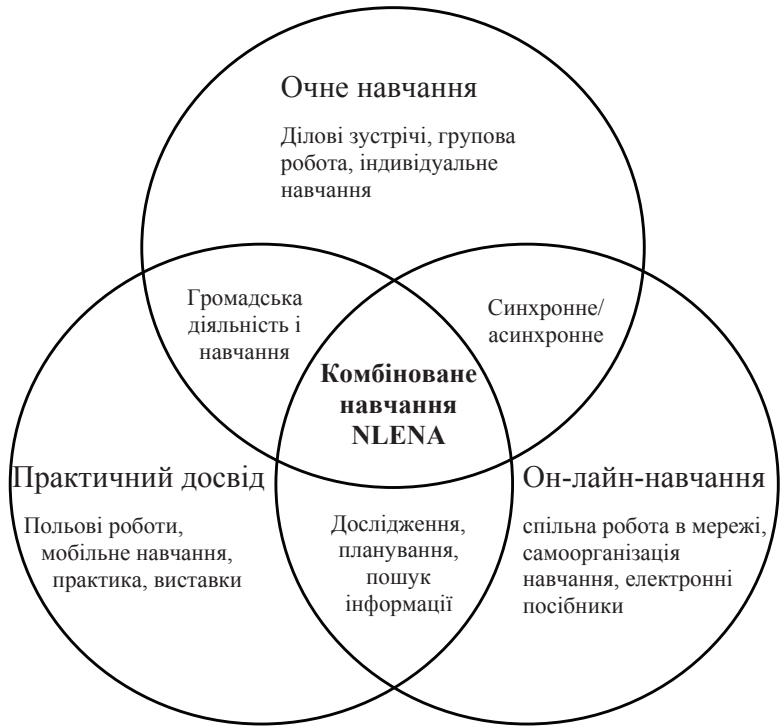

**Рис. 2. Модель інтеграції форм організації навчання в системі NLENA**

Дж. Берсін запропонував п'ять моделей комбінованого навчання (табл. 1). Перша та третя моделі Дж. Берсіна відповідають визначенню дистанційного навчання, тому що не містять елементів аудиторного навчання і розрізняються ступенем контролю тьютора за перебігом навчання. Незважаючи на те, що мультимедійність навчального середовища підкреслюється лише у першій моделі, воно притаманне й усім іншим моделям. Друга модель Дж. Берсіна відповідає нашому тлумаченню комбінованого навчання, четверта – тлумаченню мобільного тренінгу за [17, 120], а п'ята – моделі комбінованого навчання корпорації Sealund [11].

На думку Дж. Берсіна, перевіреними є такі дві моделі:

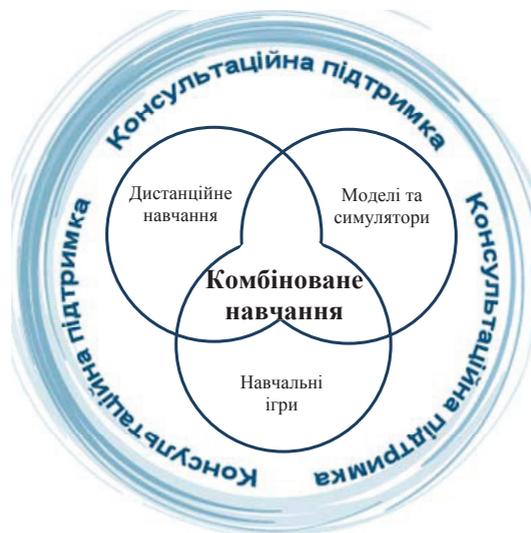

**Рис. 3. Комбіноване навчання корпорації Sealund**

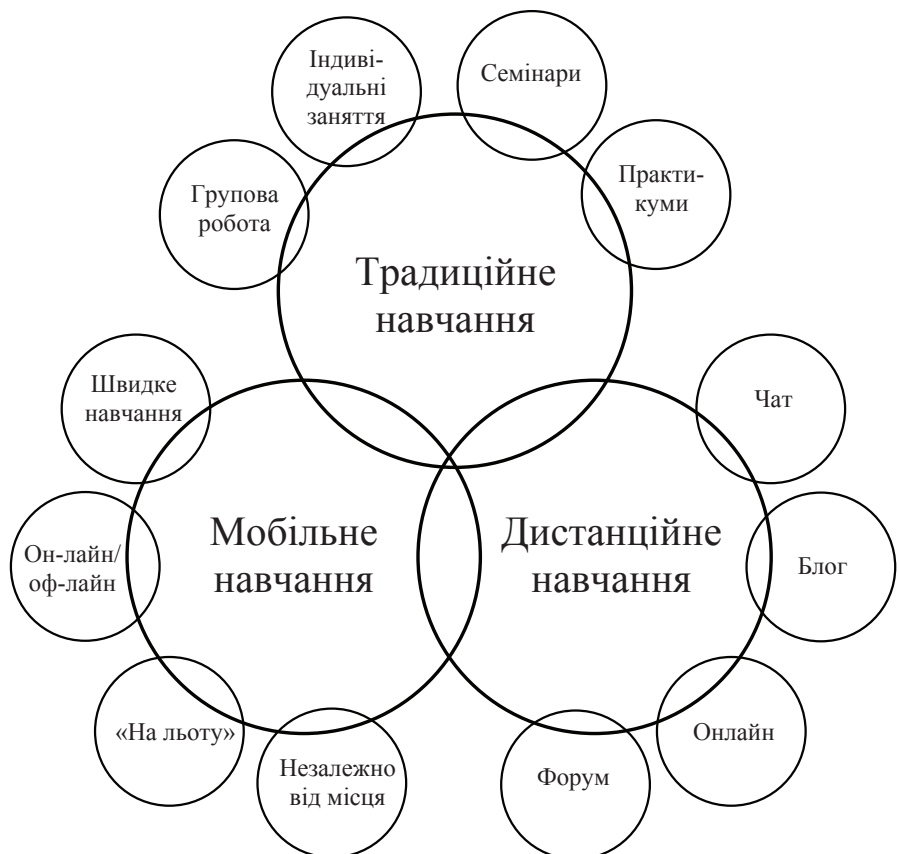

**Рис. 4. Модель комбінованого навчання за Allconsulting GmbH**

1) доповнювальна («Program Flow Model»), за якої частина традиційних форм замінюється самостійною роботою, що підтримується засобами ІКТ;

2) двостадійна («Core-and-Spoke Model»), за якої весь курс розбивається на мале ядро (вивчається за традиційними технологіями) та додаткові відомості (вивчаються за традиційними та інноваційними технологіями).

*Таблиця 1*

**Класифікація моделей комбінованого навчання**

| № | Модель | Характерні риси моделі |
|---|--------|------------------------|
| 1 | *Самонавчання у системі електронного навчання* з використанням інших комбінованих середовищ | Дистанційне навчання, за якого суб'єкт навчання занурюється у мультимедіа-середовище |
| 2 | *Навчання під керівництвом викладача*, комбіноване з самонавчанням у системі електронного навчання | За такої моделі електронне навчання підтримує традиційне аудиторне, застосовуючись для підготовки до заняття, під час заняття та після заняття |
| 3 | *Синхронне електронне навчання*, комбіноване з іншими середовищами | Основними засобами навчання стають синхронні засоби (вебінари тощо), підтримувані самонавчанням |
| 4 | *Навчання на робочому місці* | Провідною формою стає тренінг під керівництвом виробничого наставника; використовується переважно для програм формування складних умінь та навичок |
| 5 | *Орієнтована на моделювання та лабораторні роботи* | Найчастіше використовується в галузі інформаційних технологій та тренінгах, в яких може бути змодельоване необхідне середовище |

До критеріїв вибору моделі комбінованого навчання Дж. Берсін [3, 265–267] відносить: 1) тип навчального курсу (ознайомлювальний, практично-орієнтований, завершальний тощо); 2) культурні цілі (вплив очної частини курсу на досягнення цілей навчання); 3) аудиторію (розмір, розподіл навчальних ролей, рівень освіти, володіння засобами ІКТ, мотивація тощо); 4) бюджет; 5) ресурси; 6) розподіл навчального часу; 7) зміст навчання (рівень складності та інтерактивності); 8) технологічні обмеження (пропускна здатність, необхідність встановлення доповнень, відстеження діяльності, забезпечення безпеки тощо).

Б. Тунхікорн [12] запропонував модель комбінованого навчання студентських груп на основі сайту (рис. 5).

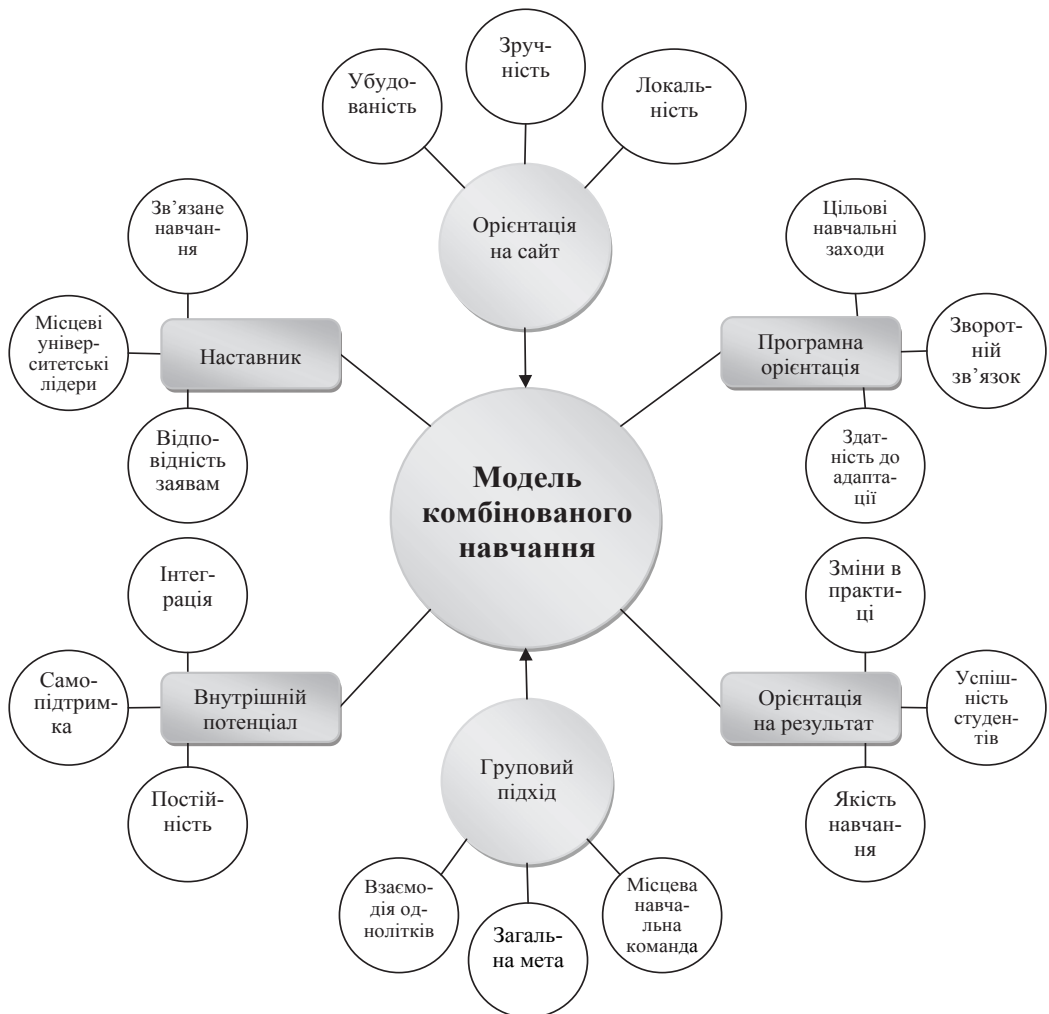

**Рис. 5. Модель комбінованого навчання за Б. Тунхікорном**

П. Джонс та інші автори [8] відповідно до ґрунтовно розробленого В.Ю. Биковим середовищного підходу до побудови моделей організаційних систем відкритої освіти [14], пропонують модель ступеневої підготовки бакалаврів. Ця модель визначає два середовища: фізичне та віртуальне, а також відповідні їм системи підтримки навчання. У віртуальному середовищі використовується обговорений у [21] поділ комунікаційних механізмів на синхронні та асинхронні. На відміну від попередніх моделей, ця модель включає одного із суб'єктів навчання – студента. Відповідно до пропозицій авторів [8] про те, що вибудувана ними модель може бути модифікована для ефективного управління комбінованим навчанням, нами було введено до неї, крім бакалаврів, ще й магістрів та докторів філософії (рис. 6).

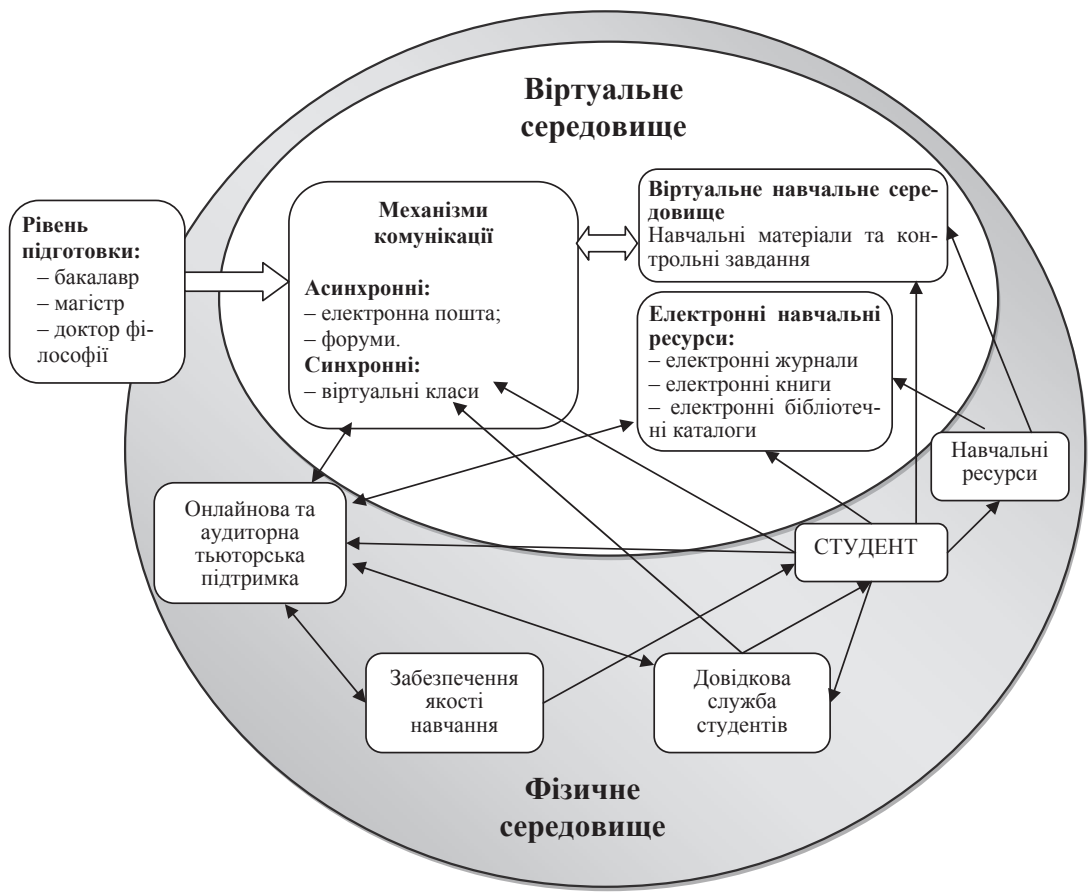

**Рис. 6. Модель комбінованого навчання в умовах ступеневої підготовки**

Модель дослідників з Університету Іогана Кеплера (м. Лінц, Німеччина) є досить спрощеною (рис. 7), проте, на відміну від попередньої, у ній присутні обидві групи суб'єктів процесу навчання – викладачі та студенти, вона відповідає уточненому тлумаченню та має ознаки педагогічної системи [6]. Зворотні стрілки у верхній частині моделі відповідають міжособистісній взаємодії суб'єктів навчання, у нижній – системному зв'язку складових технологій навчання (методів, засобів та форм організації навчання).

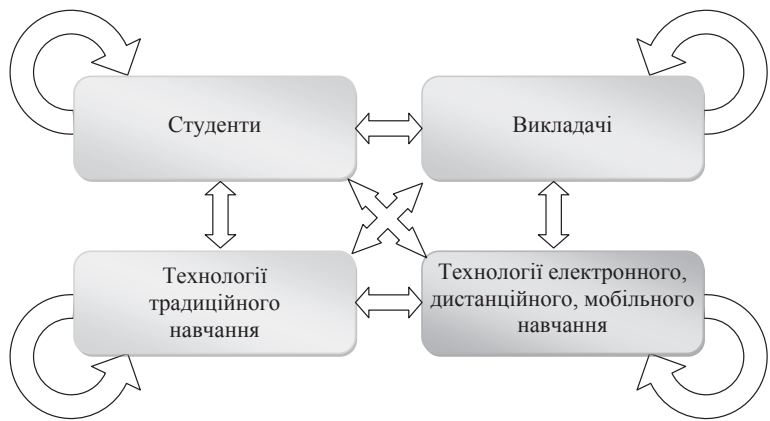

**Рис. 7. Модель комбінованого навчання як педагогічної системи**

Н.В. Рашевською [16, 85] розроблено модель, що передбачає: 1) встановлення взаємозв'язків між студентами та викладачами традиційними засобами у освітньому просторі ВНЗ та засобами мобільних ІКТ у єдиному інформаційному просторі системи освіти; 2) появу нових форм організації змішаного навчання через взаємопроникнення та інтеграцію традиційних та інноваційних форм організації відкритої освіти; 3) комбінування різних методів навчання відповідно до контексту навчання (місця, часу, стану суб'єктів навчання).

Ця модель відповідає вимогам, що передбачає модель системи відкритої освіти: зовнішньою оболонкою моделі є відкрите освітнє середовище (єдиний інформаційний простір системи освіти), внутрішньою – відкрита технологія комбінованого навчання (рис. 8).

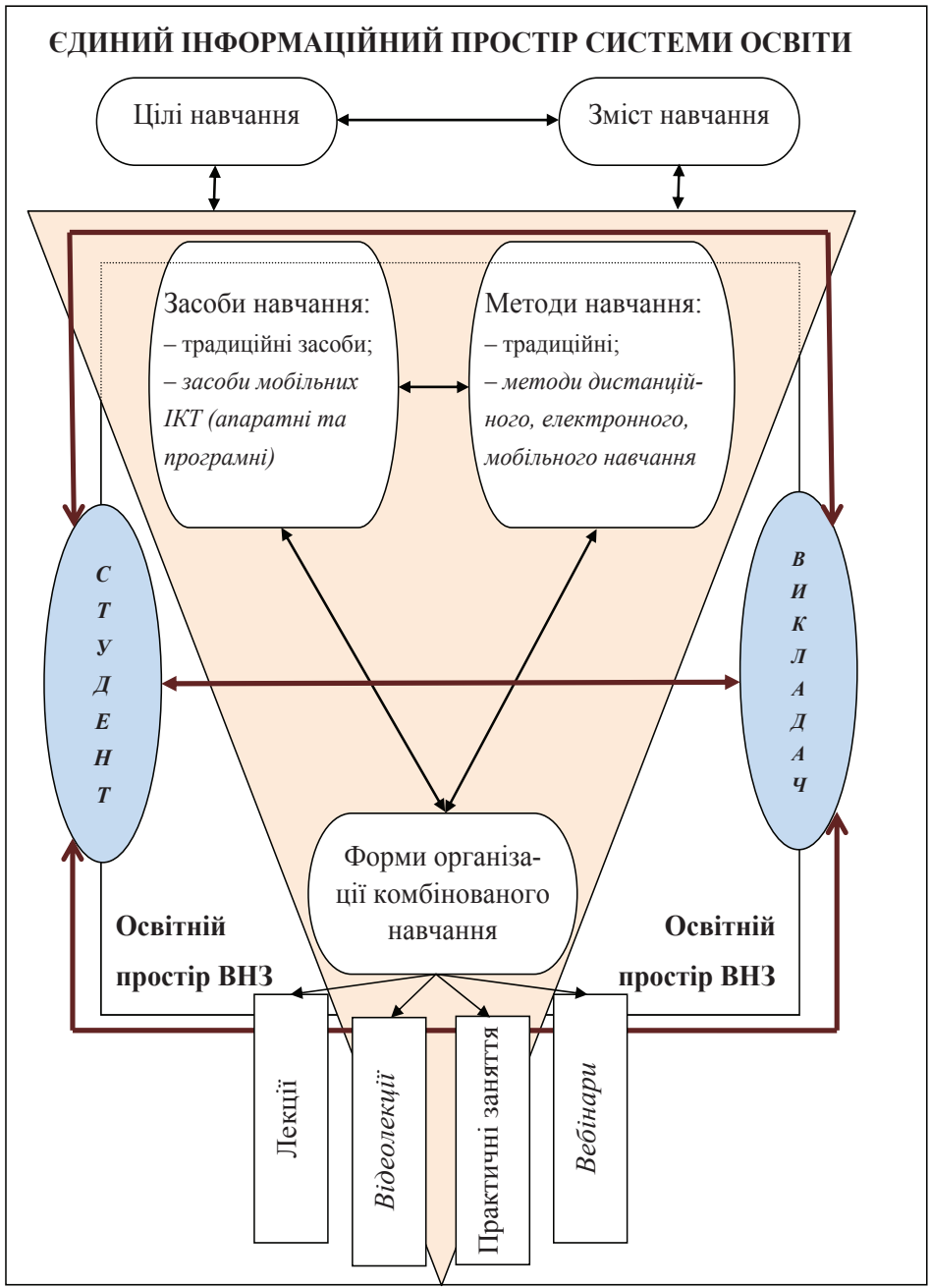

**Рис. 8. Модель комбінованого навчання у вищій школі (за Н.В. Рашевською [16])**

Провідними засобами навчання у моделі Н.В. Рашевської є засоби мобільних ІКТ (апаратні та програмні), провідними методами – методи дистанційного, електронного, мобільного навчання. Системність моделі простежується як на рівні внутрішньої оболонки (технологія комбінованого навчання як складова певної методичної системи), так і на рівні зовнішньої (утворена методична система є пов'язаною із суб'єктами процесу навчання).

У моделі Т. О'Каллаган, на відміну від попередніх, виділено рівні інтеграції форм організації комбінованого навчання засобами системи підтримки навчання (рис. 9), а у моделі організації «Education for Well-being» визначено рівні зв'язків між суб'єктами навчального процесу та їх ієрархію (рис. 10).

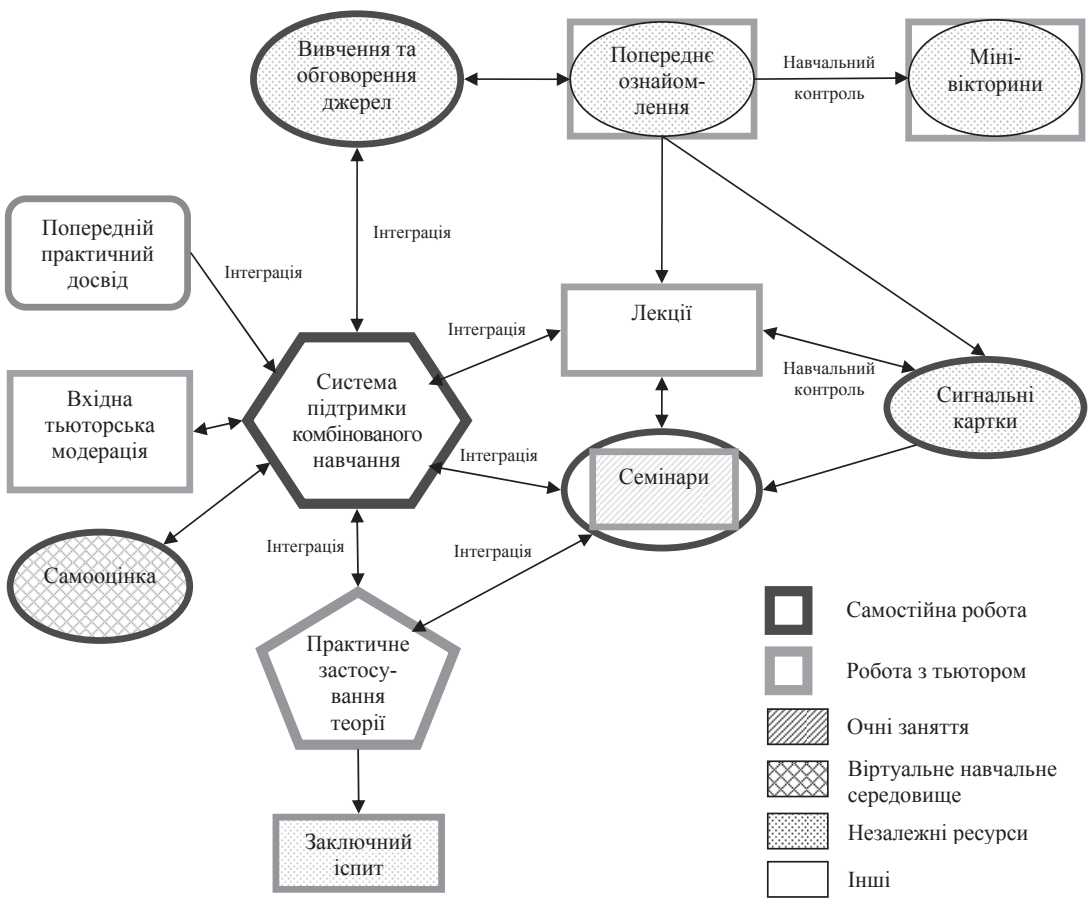

**Рис. 9. Модель комбінованого навчання з використанням системи підтримки навчання (за Т. О'Каллаган [10])**

Критерії вибору моделі комбінованого навчання, запропоновані **Дж. Берсіном**, ураховують специфіку конкретного курсу, тоді як для побудови системи комбінованого навчання у вітчизняних ВНЗ необхідним є урахування:

1) особливостей навчання не однієї дисципліни, а групи споріднених дисциплін;
2) системної та середовищної природи комбінованого навчання;
3) організаційної структури навчальної установи та її впливу на освітнє середовище:

– навчання як у мобільних (ситуативних, предметно- та практико орієнтованих) групах, так й у групах із фіксованим складом;

– наступність та ступеневість не лише у процесі навчання у ВНЗ, а й у системі «школа – коледж – університет»;

– безпосереднє відображення курикулуму у навчальному розкладі.

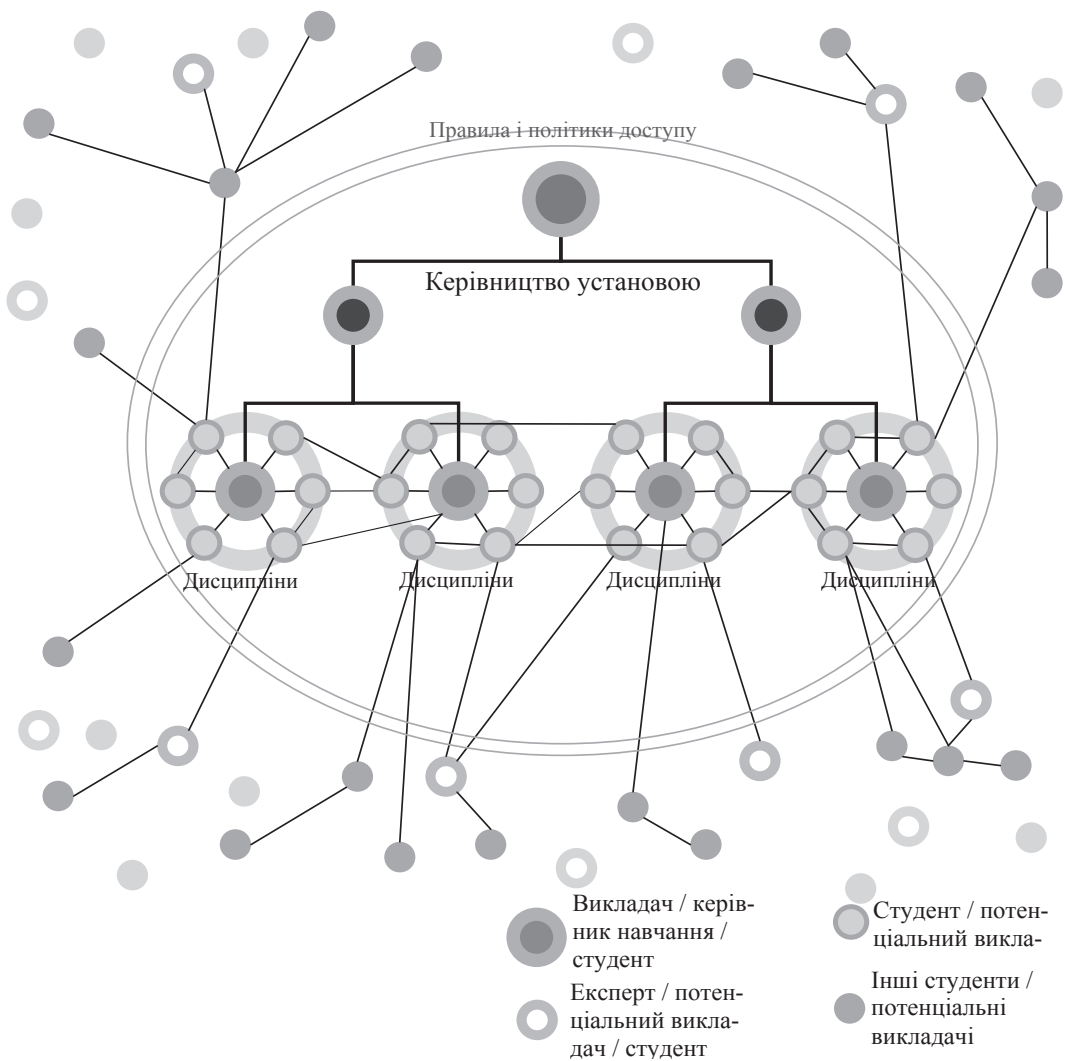

**Рис. 10. Модель комбінованого навчання з використанням системи підтримки навчання (за Education for Well-being [7])**

На жаль, жодна з розглянутих моделей не відповідає повною мірою вказаним вимогам, що зумовлює необхідність розробки нової моделі.

На рис. 11 наведено організаційну модель комбінованого навчання, **розроблену з ура**хуванням особливостей навчального процесу у вітчизняних ВНЗ. У запропонованій моделі суспільством визначаються цілі вищої освіти, які конкретизуються у галузевих стандартах вищої освіти та реалізуються закладами управління освіти, зовнішніми до освітнього середовища ВНЗ. Система управління вищою освітою безпосередньо впливає на адміністративну ієрархію освітнього середовища ВНЗ: ректорат → деканат, ректорат (деканат) → кафедра, деканат (кафедра) → студентська група. Викладачі і студенти можуть розглядатися як «пов'язані» елементи адміністративної ієрархії (фіксовані межами кафедри та групи) та як «вільні» елементи всіх вкладених середовищ.

Галузеві стандарти вищої освіти конкретизуються у навчальні плани, відображені у розкладі занять. На рівні конкретної навчальної дисципліни вони визначають цілі та зміст навчання, що разом із технологією навчання утворюють методичну систему навчання, що функціонує як у освітньому середовищі ВНЗ (на етапі її впровадження та експлуатації), так і за його межами (на етапі розробки та модифікації).

Взаємодія суб'єктів навчання відбувається як безпосередньо, так і опосередковано: через адміністративну ієрархію освітнього середовища ВНЗ та через технології навчання, що складаються з методів, форм організації та засобів навчання.

Центральною складовою запропонованої моделі є система управління навчанням, яка, з одного боку, виступає одним із засобів навчання, а з іншого – ядром, що інтегрує всі підсистеми системи комбінованого навчання у ВНЗ.

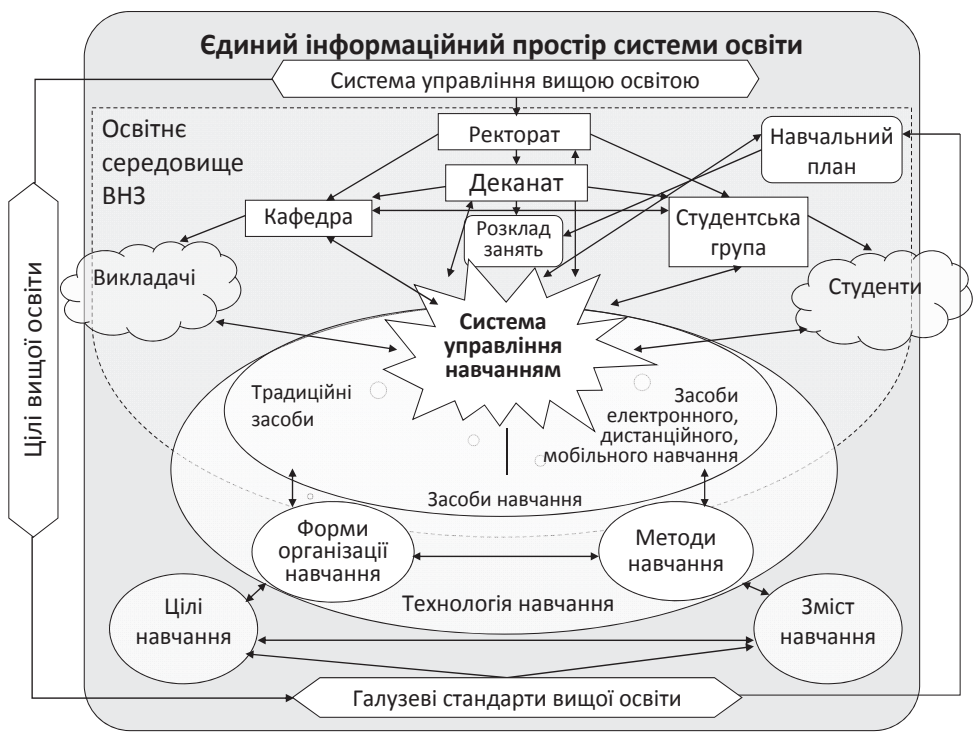

**Рис. 11. Організаційна модель комбінованого навчання у ВНЗ, що передбачає використання системи управління навчанням**

**Висновки.** Запропонована модель відповідає уточненому визначенню комбінованого навчання та відображає поточний стан розвитку теорії та методики використання ІКТ в освіті. Як зазначає Ю.В. Триус, доцільним є її фундаменталізація через заміну засобів електронного, дистанційного та мобільного навчання на інноваційні засоби ІКТ навчання, що охоплюють як існуючі класи засобів, так і ті, що будуть створені у майбутньому.

В статье на основе существующих моделей обучения представлена авторская организационная модель комбинированного обучения в ВНЗ, которая предполагает использование системы управления обучением и отображает существующее состояние развития теории и методики использования ИКТ в образовании.

*Ключевые слова: модель, комбинированное обучение, организационная модель комбинированного обучения.*

The article presents the authors' organizational model of blended learning on the basis of existing models of learning at higher educational establishments. The model provides for using the learning management system and reflects current developments of ICT use theory and methodology.

*Key words: model, blended learning, organizational model of blended learning.*